\documentclass[aps,amssymb,amsfonts,amsmath,preprint,prd]{revtex4-1}  %use this later on
\usepackage{mathrsfs}
\usepackage{graphicx}
\usepackage{flushend}
\usepackage{subcaption}
\usepackage{caption}
\usepackage{lipsum}
\usepackage[colorlinks, linkcolor=blue, anchorcolor=blue, citecolor=blue, urlcolor=blue]{hyperref}
\usepackage{tikz}
\usepackage{orcidlink}
\usepackage{booktabs}
\begin{document}
%\preprint{APS/123-QED}

\title{Gravitational spin Hall effect of electrons in Schwarzschild metric}
\author{Dan-Dan Lian \orcidlink{0000-0003-1342-5836}}
\email{liandd@mail.sysu.edu.cn}
\author{Wei-Si Qiu \orcidlink{0009-0004-6015-3671}}
\email{qiuws@mail2.sysu.edu.cn}
\author{Peng-Ming Zhang \orcidlink{0000-0002-1737-3845}}
\email{zhangpm5@mail.sysu.edu.cn}
\affiliation{School of Physics and Astronomy, Sun Yat-sen University, 519082 Zhuhai, China}

\begin{abstract}

  In this study, we derive the non-relativistic Hamiltonian for electrons within the Schwarzschild metric from covariant Dirac equations, using both the weak field approximation and the Foldy-Wouthuysen transformation. This Hamiltonian incorporates a gravitational spin-orbit coupling term, resulting in the gravitational spin Hall effect (SHE), which separates electrons by their spin. By solving the Schrödinger equation for these electrons, we investigate the gravitational SHE as they orbit a non-rotating gravitational source. Our findings reveal that the spin-dependent separation of electrons increases in proportion to their orbital periods, significantly improving the detectability of gravitational SHE. Specifically, for electrons in a low Earth orbit, the separation is estimated to be $3.0\times 10^{-12}\, \text{m}$ annually. These results indicate the practicality of detecting the gravitational SHE in electrons orbiting Earth, especially with prolonged orbital durations, underscoring the potential for quantum test of the Weak Equivalence Principle.

\end{abstract}

\maketitle

\section{Introduction}\label{sec.intro}

The spin Hall effect (SHE), a phenomenon where particles follow spin-dependent trajectories due to spin-orbit coupling, has been extensively studied in semiconductors and other materials since its initial exploration in 1971 \cite{1971JETPL..13..467D,DYAKONOV1971459}. This research has yielded numerous experimental findings and applications \cite{RevModPhys.87.1213,PMID:15539563,PhysRevLett.94.047204,2012NatCo...3..629A,Jungwirth2012,2010Sci...330.1801W}. Extending this concept to gravitational contexts leads to the notion of gravitational SHE. Predicted by Papapetrou and Corinaldesi in 1951 using the Mathisson-Papapetrou-Dixon (MPD) equations \cite{Mathisson1937,papapetrou51,Corinaldesi1951,Dixon:1970zza,Dixon:1970zz}, the gravitational SHE suggests that in a gravitational field, particles with different spins will exhibit differing trajectories due to spin-orbit coupling. This prediction appears to challenge the Weak Equivalence Principle (WEP), which posits that the trajectories of particles should be independent of their spin characteristics. Therefore, experimental verification of the gravitational SHE could provide novel insights into the WEP and gravitational theory.

The gravitational SHE has been investigated for photons using various approaches, including the Mathisson-Papapetrou-Dixon (MPD) equations \cite{Duval:2005ky,Hackmann:2014tga,PhysRevD.96.043517,Duval2019,Timogiannis:2023pop}, the Wentzel-Kramers-Brillouin (WKB) approximation \cite{Oancea2020,PhysRevD.105.104061}, and analyses based on the Energy-Momentum Tensor \cite{PhysRevD.105.104008,Lian:2023wvl}. For massive Dirac particles, Rudiger used the WKB approximation to derive their equations of motion from the Dirac equations \cite{Rudiger:1981uu}, while Audretsch explored their gravitational SHE \cite{Audretsch:1981wf}. Furthermore, Oancea et al. adapted these equations to include electromagnetic field effects \cite{Oancea:2022utx}. Marsot et al. investigated the spin-Hall effect of exotic photons on black hole horizons within Carroll symmetry \cite{Marsot:2022qkx}. Notably, these methodologies yield classical equations of motion, posing challenges for describing the evolution of quantum systems, such as atoms, in gravitational fields. 

In 2001, Obukhov derived the Hamiltonian for Dirac particles in a static gravitational field from the Dirac equations by using the Foldy-Wouthuysen (FW) transformation \cite{Obukhov:2000ih,Nicolaevici:2002zz}. Subsequently, this Hamiltonian was extended by Obukhov, Silenko, and Teryaev to accommodate Dirac particles in arbitrary gravitational fields \cite{Silenko:2004ad,Obukhov:2009qs,Obukhov:2013zca,Obukhov:2017juw}, including gravitational spin-orbit coupling terms. A semi-classical Hamiltonian was derived from the MPD equations by Deriglazov and Ram\'\i{}rez in works \cite{Ramirez:2017pmp,Deriglazov:2018vwa}. Their work \cite{Deriglazov:2017jub} may serve as a comprehensive review on this subject. It has been demonstrated that, the trajectories of Dirac particles in a uniform gravitational field are expected to be spin-dependent \cite{PhysRevD.109.044060}. The spin-dependent trajectories of Dirac particles have also been investigated using the Ishibashi-Kawai-Kitazawa-Tsuchiya matrix model \cite{Battista:2022vvl}. All these findings suggest that Dirac particles with opposite spins may follow distinct trajectories. However, detecting this divergence requires highly spin-polarized particles. This requirement underscores the importance of studying such particles, including electrons, in controlled orbital environments around Earth, where their spins can be precisely manipulated to exhibit opposite orientations.

In this study, we employ the Schwarzschild metric to model the curved spacetime outside Earth, explicitly disregarding Earth's rotation for simplicity. Following the seminal contributions of Obukhov et al. \cite{Silenko:2004ad,Obukhov:2009qs,Obukhov:2013zca,Obukhov:2017juw}, we construct the Hamiltonian for spin-polarized electrons within the Schwarzschild metric under the weak field approximation. By numerically solving the Schrödinger equations for these electrons, we obtain their wave functions and calculate the expectation values of their position operator, thereby predicting their spin-dependent separation. Our research aims to investigate this spin-dependent separation as these electrons orbit Earth and seek out the detectability of the gravitational SHE in Earth's vicinity.

Our paper is structured as follows: Section \ref{sec.Hami} introduces the FW transformation for deriving the Hamiltonian, following the approach of Obukhov et al. \cite{Obukhov:2013zca}, with a specific focus on the Hamiltonian formulation for electrons in the Schwarzschild metric. Section \ref{sec.sep} is dedicated to calculating the expectation values of the position operator to examine the orbital motion of electrons within this metric, aiming to investigate the gravitational SHE by comparing the trajectories of electrons with different spins. Section \ref{sec.class} delves into a semi-classical approximation of electron dynamics to explore the gravitational SHE as predicted by this approximation, thereby broadening the scope of our analysis. Section \ref{sec.Di} provides discussion and insights into our findings.

Our notations follow those established in the work by Obukhov et al.~\cite{Obukhov:2013zca}. We use Latin letters \(i,j,k,\ldots = 0,1,2,3\) to denote the world indices of tensorial objects, while the initial letters of the Greek alphabet, \(\alpha,\beta,\gamma,\ldots = 0,1,2,3\), represent the tetrad indices. Spatial indices for three-dimensional constructs are indicated by the early Latin letters \(a,b,c,\ldots = 1,2,3\). We distinguish specific tetrad indices with hats. Additionally, we normalize units such that \(\hbar = c = m_e = 1\), where \(c\) is the speed of light and \(m_e\) denotes the electron mass, to simplify expressions. The metric signature has been adopted to $\eta_{\alpha\beta}=\text{diag}(+,-,-,-)$.

\section{The Hamiltonian of electrons in Schwarzschild metric}\label{sec.Hami}

In this section, we follow the results given in the work \cite{Obukhov:2013zca} and provide a succinct overview of the derivation of the Hamiltonian for electrons within the Schwarzschild metric under the weak field approximation. The general form of the line element of an arbitrary gravitational field is expressed as
\begin{equation}\label{metric}
ds^2=V^2 dt^2-\delta_{\hat{a}\hat{b}}W^{\hat{a}}_{c}W^{\hat{b}}_{d}(dx^c-K^cdt)(dx^d-K^d dt).
\end{equation}
Specifically, the Schwarzschild metric in harmonic coordinates can be given by:
\begin{equation}
  \text{d}s^2=\left(\frac{1-GM/r}{1+GM/r}\right)\text{d}t^2 - \left(1+\frac{GM}{r}\right)^2 (\text{d}\vec{x})^2-\left(\frac{1+GM/r}{1-GM/r}\right)\frac{G^2 M^2}{r^4}(\vec{x}\cdot \text{d}\vec{x})^2.
\end{equation}
Under the weak field approximation ($GM/r\ll 1$), its line element  is simplified as:
\begin{equation}\label{LT}
  ds^2 = (1-\frac{2GM}{r})dt^2-(1+\frac{2GM}{r})(dx^2+dy^2+dz^2),
\end{equation}
where terms of order higher than the first in $GM/r$ are neglected, $M$ represents the mass of gravitational source, and $r\equiv\sqrt{x^2+y^2+z^2}$.  At the first order in $GM/r$, the quantities $V$, $W^{\hat{a}}_c$ and $K^c$ take the forms 
\begin{equation}\label{VW}
  V\simeq 1-\frac{GM}{r},\quad W^{\hat{a}}_c = W\delta^{\hat{a}}_c,\quad W\simeq 1+\frac{GM}{r},\quad \vec{K}\equiv \{ K^c\} = 0.
\end{equation}

It is important to note that in this analysis, the electromagnetic field of electrons are disregarded. Consequently, the equation of motion in a curved spacetime is governed by the following covariant Dirac equation
\begin{equation}\label{eqdirac}
   (i \gamma^\alpha D_\alpha - m_e)\Psi = 0,\quad \alpha = 0,1,2,3;
\end{equation}
where $D_\alpha\equiv e^i_\alpha D_i$ with $D_i = \partial_i + \frac{i}{4}\sigma^{\alpha\beta}\Gamma_{i\alpha\beta}$ denotes the spinor covariant derivatives, $\Gamma_i^{\ \alpha\beta}=-\Gamma_i^{\ \beta\alpha}$ representes the Lorentz connection, and $\sigma^{\alpha\beta} = \frac{i}{2}(\gamma^\alpha\gamma^\beta-\gamma^\beta\gamma^\alpha)$. In the Schwarzschild metric, under the Schwinger gauge ($e^{\hat{0}}_a = 0$), the tetrad components are
\begin{equation}\label{tetrad}
  e^{\hat{0}}_i=V\delta^{\hat{0}}_i,\quad e^{\hat{a}}_i\simeq W\delta^a_i,\quad e^i_{\hat{0}} \simeq (1+\frac{GM}{r})\delta^i_0 , \quad e^i_{\hat{a}}\simeq (1-\frac{GM}{r})\delta^i_{\hat{a}}.
\end{equation}

Consequently, the Dirac equation can be reformulated in the Schrödinger form
\begin{equation}\label{eqsch}
   i\partial_t\psi = \mathcal{H}\psi,
\end{equation}
where $\psi \equiv (\sqrt{g}e^0_{\hat{0}})^{1/2}\Psi$ denotes the rescaled wave function, ensuring the Hamiltonian $\mathcal{H}$ is Hermitian. The Hamiltonian is given by
\begin{equation}\label{eqham}
   \mathcal{H} = \beta m_e V + \frac{1}{2}(p_b\mathcal{F}^b_{\ a}\alpha^a+\alpha^a\mathcal{F}^b_{\ a}p_b).
\end{equation} 
Here, $\vec{p}\equiv \{p_a\}$ represents the momentum operator, with $p_a = -i\partial_a$, and $\beta=\gamma^{\hat{0}}$. The notations $\alpha^a$, $\vec{\Sigma}\equiv\{\Sigma^a\}$ and $\mathcal{F}^a_{\ b}$ are defined as
\begin{equation}
  \alpha^a = \gamma^{\hat{0}}\gamma^a,\quad \Sigma^1=i\gamma^{\hat{2}}\gamma^{\hat{3}},\quad \Sigma^2=i\gamma^{\hat{3}}\gamma^{\hat{1}},\quad \Sigma^3=i\gamma^{\hat{1}}\gamma^{\hat{2}}, \quad \mathcal{F}^a_{\ b}=VW^a_{\ \hat{b}}.
\end{equation}
Under the weak field approximation, $\mathcal{F}^a_{\ b}$ is approximated as
\begin{equation}
  \mathcal{F}^a_{\ b}\simeq (1-\frac{2GM}{r})\delta^a_b.
\end{equation}

In the work of Obukhov et al. \cite{Obukhov:2013zca}, the Hamiltonian $\mathcal{H}$, transformed using the FW transformation within the Schwarzschild metric, is presented as:
\begin{equation}
  \mathcal{H}_{FW}=\beta\epsilon' + \frac{1}{16}\left\{\frac{1}{\epsilon'},2\epsilon^{cae}\Pi_e \{p_b, \mathcal{F}^d_{\ c}\partial_d\mathcal{F}^b_{\ d} \}\right\} + \frac{m_e}{4}\epsilon^{cae}\left\{ \frac{1}{\mathcal{T}},\{p_b,\mathcal{F}^d_{\ c}\mathcal{F}^b_{\ a}\partial_b V\}\right\},
\end{equation}
where $\Pi^a=\beta\Sigma^a$ denotes the polarization operator, and $\epsilon_{cae}$ is the Levi-Civita symbol with $\epsilon_{123}=1$. The expressions for $\epsilon'$ and $\mathcal{T}$ are given by
\begin{equation}
  \epsilon'=\sqrt{m^2_e V^2+\frac{1}{4}\delta^{ac}\{p_b,\mathcal{F}^b_{\ a}\}\{ p_d,\mathcal{F}^d_{\ c}\}},\quad \mathcal{T}=2\epsilon'^2+\{\epsilon',m_e V\}.
\end{equation}
In these equations, the reduced Planck's constant $\hbar$ and the speed of light $c$ have been normalized to unity ($\hbar=c=1$).

Under the weak field approximation, $\mathcal{H}_{FW}$ is approximated as
\begin{align}\label{hami}
  \mathcal{H}_{FW} &\simeq \beta m_e \left(1-\frac{GM}{r}+\frac{p^2}{2m^2_e}-\frac{3GMp^2}{2m^2_e r}-\frac{3iGM}{2r^3 m^2_e}\vec{r}\cdot\vec{p}\right) + \frac{3GM\beta}{4m_e r^3}\vec{\Sigma}\cdot\vec{l},
\end{align}
where $\vec{l}=\vec{r}\times\vec{p}$ denotes the orbital angular momentum operator of electrons. In the analysis of electrons orbiting the gravitational source, the gravitational potential is significantly less than unity $GM/r\ll 1$, and the average momentum $\langle \vec{p}\rangle$ is substantially smaller than the electron mass $m_e=1$.  Consequently, in the FW Hamiltonian $\mathcal{H}_{FW}$, terms of order higher than the first in $GM/r$ and $p^2$ are neglected. Moreover, only the first-order terms of the commutator $[p_a,GM/r]$ are considered. Under these approximations, our Hamiltonian aligns with that presented in Obukhov et al.~\cite{Obukhov:2009qs}, while there are subtle differences from the semi-classical Hamiltonian described by Ram\'\i{}rez et al.\cite{Ramirez:2017pmp}.

The second term in Eq. \eqref{hami} is directly proportional to $\vec{\Sigma} \cdot \vec{l}$, signifying the interaction between the electron's spin and its orbital angular momentum. Specifically, when an electron's spin is orthogonal to its orbital angular momentum, the gravitational SHE induced by this term  is expected to vanish. Considering the above discussions, this investigation primarily explores the gravitational SHE of electrons with spin polarization along the $z$-axis, moving within the $x-y$ plane. In this case, the gravitational SHE, induced by the second term, is expected to reach its maximum. 

\section{gravitational spin Hall effect of electrons}\label{sec.sep}

As previously discussed, the average momentum $\langle \vec{p}\rangle$ of electrons considered in this study is significantly smaller than their mass $m_e$, justifying the non-relativistic approximation. Within this regime, the negative energy components of the Dirac wave functions are negligibly small compared to the positive energy components. Consequently, the FW Hamiltonian can be further simplified to the following form:
\begin{align}\label{hami-nr}
  \mathcal{H}_{FW}&\simeq \left(-\frac{GM}{r}+\frac{p^2}{2}-\frac{3GMp^2}{2r}-\frac{3iGM}{2r^3}\vec{r}\cdot\vec{p}\right) + \frac{3GM}{4r^3}\vec{\sigma}\cdot\vec{l},
\end{align}
where $\vec{\sigma}=\{\sigma^a\}$ represents the Pauli matrices, and the mass of electrons has been taken to $m_e=1$. The wave function of electrons is then governed by the equation:
\begin{equation}\label{non-sch}
  \partial_t\psi (\vec{x},t)=\mathcal{H}_{FW}\psi (\vec{x},t).
\end{equation}

In our analysis, we adopt the Dirac representation while disregarding these negative energy components, thereby describing an electron with a two-component wave function. Throughout this study, the initial states of electrons in momentum space are chosen as follows \cite{Karlovets:2018iww}:
\begin{equation}\label{init}
  \psi(\vec{p}) = \left(\frac{2\sqrt{\pi}}{\sigma}\right)^{3/2}\omega\exp\left(-\frac{(\vec{p}-\vec{p}_0)^2}{2\sigma^2}\right),
\end{equation}
where $\omega$ denotes the eigenstate of the Pauli matrix $\sigma^3$ with an eigenvalue $\lambda$. Specifically, $\omega=(1,0)^T$ corresponds to the eigenvalue $\lambda=+1$, and $\omega=(0,1)^T$ to $\lambda=-1$.

The methodology for solving the Schrödinger equation (Eq.~\ref{non-sch}) presented in this study is consistent with the approach outlined by Lian et al.~\cite{Lian:2023wvl}, and is executed in three systematic steps. Initially, the equation undergoes a Fourier Transform $\mathscr{F}$ within three-dimensional coordinate space, defined as:
\begin{equation}
  \mathscr{F}(f(\vec{x},t))\equiv \frac{1}{(2\pi)^{3/2}}\int f(\vec{x},t)\exp (-i \vec{p}\cdot \vec{x}) \text{d}^3x.
\end{equation}
This transformation effectively recasts the equations from a set of second-order partial differential equations into a set of first-order ordinary differential equations in momentum space. 

Subsequently, the midpoint method is adopted to solve the transformed equations in momentum space, starting from the initial conditions specified in Eq.~\ref{init}. This approach facilitates the derivation of the electron's wave function in momentum space, $\psi(\vec{p},t)$. The final step computes the wave function in coordinate space, $\psi(\vec{x},t)$, through an inverse Fourier Transform $\mathscr{F}^{-1}$, expressed as:
\begin{equation}
  \psi (\vec{x},t)= \mathscr{F}^{-1}(\psi (\vec{p},t))\equiv \frac{1}{(2\pi)^{3/2}}\int \psi (\vec{p},t)\exp (i \vec{p}\cdot \vec{x}) \text{d}^3p.
\end{equation} 

\subsection{Difference in trajectories of electrons with opposite spins}\label{subsec.diffe}
\begin{figure}[htb]
  \centering
  \begin{subcaptionblock}{.42\textwidth}
    \centering
    \includegraphics[scale=0.46]{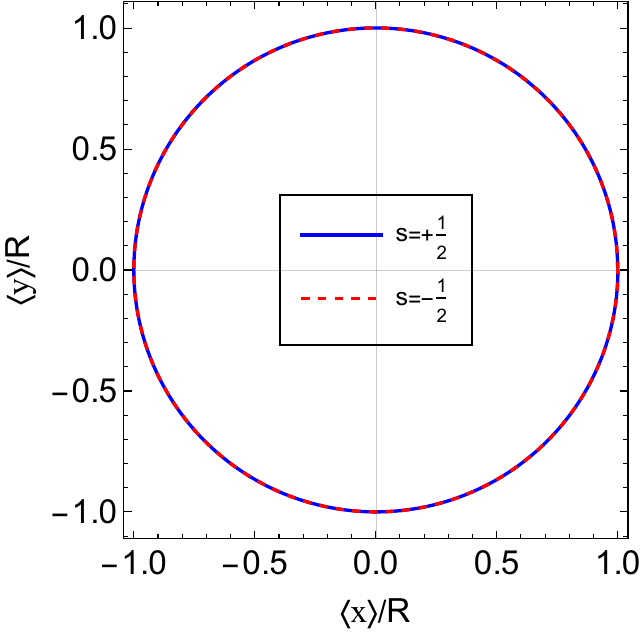}
    \caption{Trajectories of Electrons}
  \end{subcaptionblock}
  \begin{subcaptionblock}{.48\textwidth}
      \centering
      \includegraphics[scale=0.44]{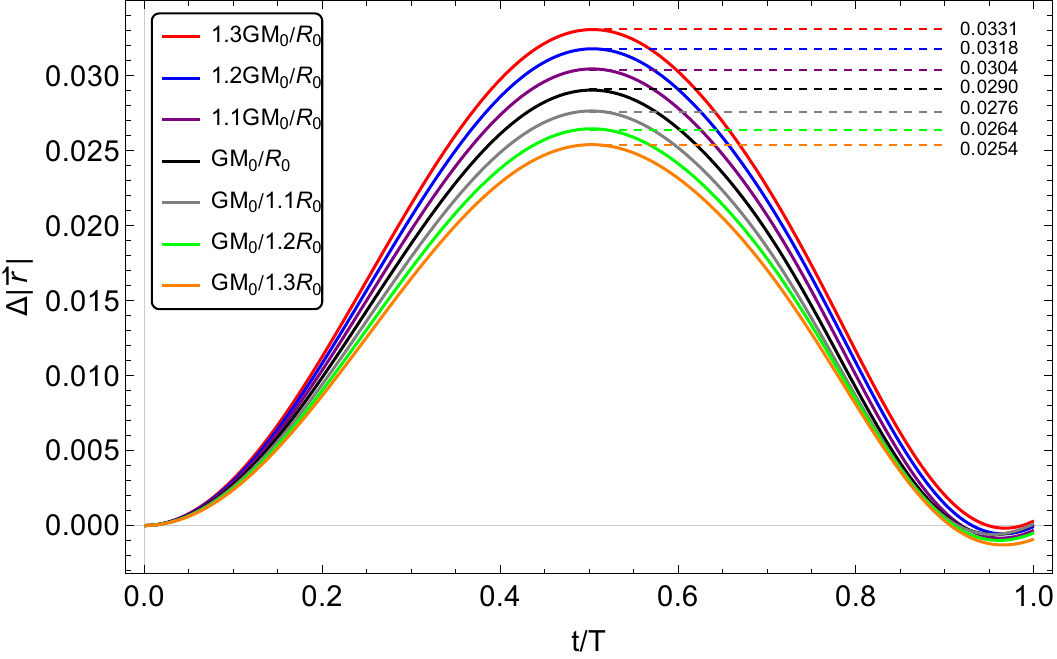}
      \caption{Difference in radial distances of electrons}
  \end{subcaptionblock}
\caption{Trajectories of two electrons with opposite spins $s=\pm 1/2$. The right panel shows the difference in their radial distances from the gravitational source, defined as $\Delta |\vec{r}| = |\vec{r}_{1/2}| - |\vec{r}_{-1/2}|$. For these demonstrations, the initial positions and momenta of the electrons are set to $\vec{r}_0 = (R, 0, 0)$ and $\vec{p}_0 = (0, \sqrt{GM/R}, 0)$, respectively, with parameters $R_0=10^5$ and $GM_0=1$. Here, $T\simeq 2\pi R/|\vec{p}_0|$ denotes the orbital period.}\label{fig1}
\end{figure}

Once the wave function $\psi(\vec{x},t)$ is determined numerically, the position of this electron can be straightforwardly computed by 
\begin{equation}
\langle x^a\rangle = \langle \psi|x^a|\psi\rangle.
\end{equation}
The trajectories of two electrons with opposite spins, $s=\pm 1/2$,  are depicted in Figures \ref{fig1} . Their initial conditions are uniformly set with the momentum $\vec{p}_0=(0,\sqrt{GM/R},0)$ and position $\vec{r}_0=(R,0,0)$, where $R$  denotes the initial distance of the electrons from the gravitational source. 

As illustrated in Fig.~\ref{fig1}, two electrons with opposite spins $s=\pm 1/2$, orbit the gravitational source in the $x-y$ plane along distinct trajectories. The results suggest that, although the electrons approximately maintain their individual closed orbits, their spatial trajectories do not precisely overlap. The right panel highlights the difference in their radial distances from the gravitational source, denoted as $\Delta |\vec{r}| = |\vec{r}_{1/2}| - |\vec{r}_{-1/2}|$. This difference peaks near $t \simeq T/2$ and gradually diminishes, nearing zero as $t$ approaches $T$, where $T\simeq 2\pi R/|\vec{p}_0|$ represents the orbital period. 

Notably, $\Delta |\vec{r}|$ does not precisely vanish  at $t=T$. Towards the end of this specific orbital period, the numerical results exhibit complexity. For example, when $t/T > 0.8$, the blue and green curves, representing different gravitational potentials ($GM/R=1.2GM_0/R_0$ and $GM/R=GM_0/1.2R_0$, respectively), converge closely. This behavior suggests that, $\Delta |\vec{r}|$ can be expected to tend towards zero at $t=T$. Surprisingly, the deviations from this expected trend show little sensitivity to the gravitational potential $GM/R$, suggesting that numerical errors rather than physical processes might drive these deviations. 

Ignoring the numerical inaccuracies, the behavior of $\Delta |\vec{r}|$, which peaks near $t \simeq T/2$ and decreases to nearly zero as $t$ approaches $T$, aligns with the anticipated behavior of radial distance difference for two electrons moving along elliptical orbits with different eccentricities. This alignment indicates that the trajectory of a spin-polarized electron deviates slightly from the circular orbit, approximating an elliptical orbit instead.

Additionally, the peak value of this difference, denoted as $\Delta |\vec{r}|_\text{max}$, is influenced by the mass of the gravitational source, scaling with $\sqrt{GM}$, and inversely with the square root of the initial distance from the source, $\sqrt{R}$, as described by the equation
\begin{equation}\label{num-rad}
  \Delta |\vec{r}|_{\text{max}} \propto \sqrt{\frac{GM}{R}}.
\end{equation}
This relationship indicates that, the difference in orbital eccentricities for the electrons with opposite spins increases with the gravitational source mass $M$, and decreases as the initial distance $R$ grows.

\subsection{Separation between electrons with opposite spins}

\begin{figure}[htb]
  \centering
  \includegraphics[scale=0.6]{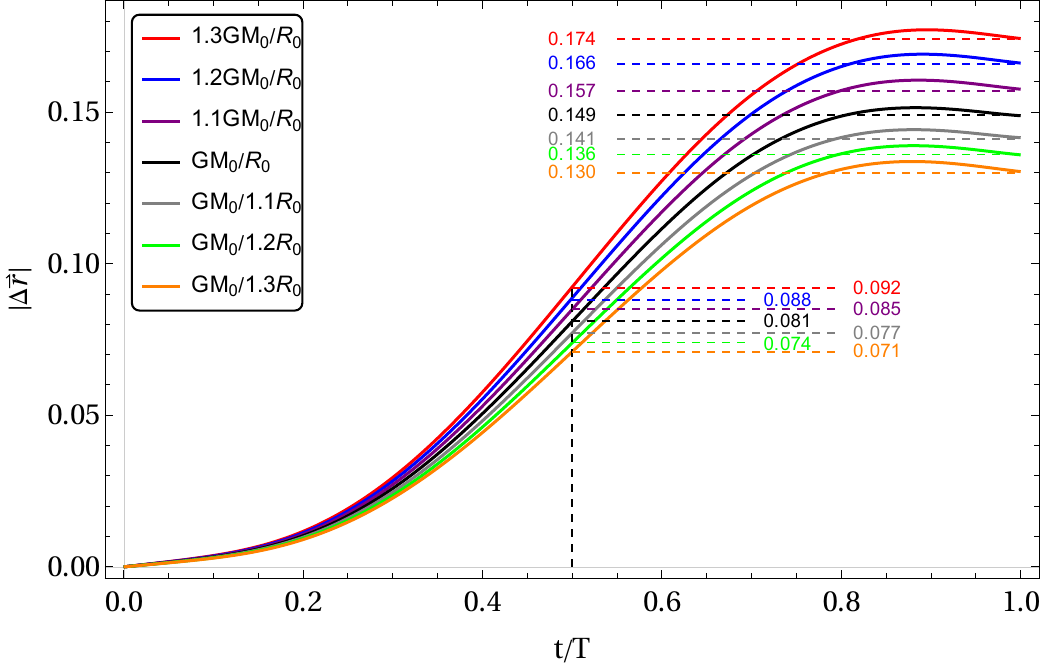}
\caption{Spatial separation $|\Delta\vec{r}| = |\vec{r}_{1/2} - \vec{r}_{-1/2}|$ of two electrons with an opposite spin $s=\pm 1/2$. The gravitational source mass has been set to $GM_0=1$, while the initial distance to the gravitational source is set as $R_0=10^5$.}\label{fig2}
\end{figure}

When these electrons orbit the gravitational source in the $x-y$ plane, their orbital angular momentum $\langle\vec{l}\rangle=\langle \psi|\vec{l}|\psi\rangle$ aligns with their spin $\vec{s}=\langle \psi|\vec{\sigma}|\psi\rangle/2$. This alignment introduces the spin-orbit coupling effect, described by $3GM\vec{\sigma}\cdot \vec{l}/4r^3$, leading to their spatial separation in three dimensions, a manifestation of the gravitational SHE. As shown in Fig.~\ref{fig2}, after an equal duration of motion $t=T/2$, their spatial separation, denoted as $|\Delta\vec{r}|=|\vec{r}_{1/2} - \vec{r}_{-1/2}|$, can also be approximately proportional to both $\sqrt{GM}$ and $\sqrt{1/R}$, as given by:
\begin{equation}\label{num-res}
 |\Delta \vec{r}|_{\text{T/2}} \propto \sqrt{\frac{GM}{R}}.
\end{equation}

In subsection \ref{subsec.diffe}, it has been shown that the electrons with opposite spins orbit the gravitational source on elliptical paths of differing eccentricities. Consequently, the orbital period of an electron with spin $s=1/2$, denoted by $T_{1/2}$, differs from that of an electron with opposite spin $s=-1/2$, represented by $T_{-1/2}$. After a period of $nT_{1/2}$, the electron with spin $s=1/2$ returns to its initial position $\vec{r}_0$, whereas the electron with spin $s=-1/2$ does not, due to the inequality $T_{1/2} \neq T_{-1/2}$. Given the minimal difference in their orbital periods, the separation between these electrons can be approximated as
\begin{equation}\label{n-orbit}
  |\Delta \vec{r}|_\text{n} \sim nv_0 |T_{1/2} - T_{-1/2}| \sim n |\Delta \vec{r}|_\text{T},
\end{equation}
where $v_0 = |\vec{v}_0| \simeq |\vec{p}_0|$ is the velocity of the electrons, and $|\Delta \vec{r}|_\text{T}$ is the separation between the two electrons after one orbital period ($n=1$). This formula suggests that the separation $|\Delta \vec{r}|$ can be expected to increase progressively over subsequent orbital periods.

Remarkably, electrons with opposite spins initiate their motion from the same position, $\vec{r}_0=(R,0,0)$, each possessing the same initial momentum, $\vec{p}_0=(0,\sqrt{GM/R},0)$. Despite their identical momenta, the observed separation between these electrons implies differing velocities. This difference in velocities indicates that the gravitational spin-orbit coupling affects the momentum-velocity relationship of electrons, making the simple relation of $\vec{p}=m_e\vec{v}$ inapplicable, even within the non-relativistic approximation.

\section{Semi-Classical approximation for electron dynamics}\label{sec.class}

In this analysis, we employ a semi-classical approximation by representing the quantum operators $\vec{p}$, $\vec{r}$, $\vec{l}$, and $\vec{\sigma}$ with their expectation values within the FW Hamiltonian. This approach derives a classical version of the Hamiltonian, expected to represent a semi-classical approximation of electron dynamics. To assess the effectiveness of the semi-classical approximation, we compare the gravitational SHE predictions derived from quantum dynamics with those obtained through this approximation.

Upon substituting the expectation values of the quantum operators into Eq. \eqref{hami-nr}, we obtain a classical Hamiltonian represented as:
\begin{equation}\label{c-hami}
H_{\text{class}}=-\frac{GM}{r} + \left(1-\frac{3GM}{r}\right)\frac{ \vec{p}^2}{2} + \frac{3GM}{2r^3}\vec{s}\cdot\vec{l},
\end{equation}
where $\vec{p}$, $\vec{s}$, and $\vec{l}$ denote the expectation values of the electron's momentum, spin ($\frac{1}{2}\vec{\sigma}$), and orbital angular momentum operators, respectively, calculated as $\langle \psi | \cdot |\psi\rangle$. The term $\frac{3iGM}{2r^3}\vec{r}\cdot\vec{p}$ is expected to disappear in the classical Hamiltonian owing to the Hermitian characteristics of $\mathcal{H}_{FW}$.

For a classical particle, its dynamics are governed by Hamilton's equations, as specified by the Hamiltonian in Eq.~\eqref{c-hami}:
\begin{equation}\label{c-eom}
\frac{dx^a}{dt} = \frac{\partial H}{\partial p_a},\quad \frac{dp_a}{dt} = -\frac{\partial H}{\partial x^a},
\end{equation}
where $\left\{x^a\right\}=\vec{r}$ and $\left\{p_a\right\}=\vec{p}$ represent the components of the position vector $\vec{r}$ and momentum $\vec{p}$, respectively. In this section, the electrons are expected to orbit within the $x-y$ plane, as indicated by their initial conditions: position $\vec{r}_0=(R,0,0)$ and momentum $\vec{p}_0=(0,\sqrt{GM/R},0)$. Additionally, the electrons' spin is aligned with the $z$-axis, perpendicular to the orbital plane. Consequently, the classical dynamics of the electrons are effectively reduced to a two-dimensional analysis, described by
\begin{equation}\label{c-reom}
\left\{
  \begin{split}
    &\frac{dx}{dt} = \left(1-\frac{3GM}{r}\right)p_x - \frac{3GM y}{2r^3}s,\\
    &\frac{dy}{dt} = \left(1-\frac{3GM}{r}\right)p_y + \frac{3GM x}{2r^3}s\\
  \end{split}\right.
\end{equation}
and 
\begin{equation}\label{c-peom}
  \left\{
    \begin{split}
      &\frac{dp_x}{dt} = -\left(1+\frac{3p^2}{2}\right)\frac{GMx}{r^3}-\frac{3GMp_y}{2r^3}s+\frac{9GM xl_z}{2r^5}s\\
      &\frac{dp_y}{dt} = -\left(1+\frac{3p^2}{2}\right)\frac{GMy}{r^3}+\frac{3GMp_x}{2r^3}s+\frac{9GM yl_z}{2r^5}s,
    \end{split}
  \right.
\end{equation}
where, $s=\pm 1/2$ signifies the electron's spin along the $z$-axis, and $l_z=xp_y-yp_x$ corresponds to the electron's orbital angular momentum in this direction. With $z=0$ and $p_z=0$, we reduce $r$ and $p^2$ to $r=\sqrt{x^2+y^2}$ and $p^2=p_x^2+p_y^2$, respectively.

As detailed in Section \ref{sec.sep}, electrons with an initial momentum of $|\vec{p}_0|=\sqrt{GM/R}$ tend to maintain an approximately circular orbit with radius $R$, when these electrons are not spin-polarized. Semi-classical dynamics, as described by Equations \eqref{c-reom} and \eqref{c-peom}, suggest that the influence of gravitational spin-orbit coupling is markedly smaller than that of the gravitational potential. Consequently, deviations in position $\Delta x^i_s=x^i_s-x^i_{s=0}$ and momentum $\Delta p_{i,s}=p_{i,s}-p_{i,s=0}$ for a spin-polarized electron, compared to an electron without spin polarization, are slight, quantified as $\Delta x^i_s/R\ll 1$ and $\Delta p_{i,s}/p\ll 1$, respectively. Here, $p=|\vec{p}_{s=0}|$ represents the magnitude of momentum for electrons without spin polarization.

For the electron without spin polarization, circularly orbiting a gravitational source with radius $R$, its position $x^i$, momentum $p_i$ and orbital angular momentum $l_z$ are approximated by
\begin{align}
x\simeq R\cos(\theta),\quad y\simeq R\sin(\theta),\quad p_x\simeq -p\sin(\theta),\quad p_y\simeq p\cos(\theta),\quad l_z\simeq R p,
\end{align}
where $\theta\simeq 2\pi t/T$ is the azimuthal angle. Taking these expressions into account and limiting the analysis to first-order deviations $\Delta x^i_s$ and $\Delta p_{i,s}$, the gravitational spin-orbit coupling effects can be approximated as
\begin{equation}
  \left\{
    \begin{split}
&\frac{d \Delta x_s}{d\theta}\simeq \frac{R\Delta p_{x,s}}{p}-\frac{3GMs}{2Rp}\sin\theta,\\
&\frac{d \Delta y_s}{d\theta}\simeq \frac{R\Delta p_{y,s}}{p}+\frac{3GMs}{2Rp}\cos\theta,
    \end{split}
    \right.
\end{equation}
and
\begin{equation}
  \left\{
  \begin{split}
  &\frac{d\Delta p_{x,s}}{d\theta}\simeq \frac{GM(1+3\cos(2\theta))\Delta x_s}{2pR^2}+\frac{3GM\sin(2\theta)\Delta y_s}{2pR^2}+\frac{3GMs\cos\theta}{R^2},\\
  &\frac{d\Delta p_{y,s}}{d\theta}\simeq \frac{GM(1-3\cos(2\theta))\Delta y_s}{2R^2 p}+\frac{3GM\sin(2\theta)\Delta x_s}{2pR^2}+\frac{3GMs\sin\theta}{R^2},
  \end{split}
  \right.
\end{equation}
where $d\theta/dt$ has been approximated as $p/R$.

At the initial time when $t=0$, we set the deviations in position and momentum to zero, denoted as $\Delta x^i_s|_{t=0}=0$ and $\Delta p_{i,s}|_{t=0}=0$, respectively. Under these conditions, the differences in positions between spin-polarized and unpolarized electrons are governed by the following equations:
\begin{equation}\label{c-sep}
  \left\{
\begin{split}
  &\frac{d^2 \Delta x_s}{d\theta^2}\simeq \frac{(1+3\cos(2\theta))\Delta x_s}{2} + \frac{3\sin(2\theta)\Delta y_s}{2} + \frac{3s}{2}\sqrt{\frac{GM}{R}}\cos\theta,\\
  &\frac{d^2 \Delta y_s}{d\theta^2}\simeq \frac{(1-3\cos(2\theta))\Delta y_s}{2} + \frac{3\sin(2\theta)\Delta x_s}{2} + \frac{3s}{2}\sqrt{\frac{GM}{R}}\sin\theta\\
\end{split}
  \right.
\end{equation}
with the initial conditions specified as:
\begin{align}\label{c-init}
  \Delta x_s|_{t=0},\quad \Delta y_s|_{t=0}=0;\quad \frac{d\Delta x_s}{d\theta}|_{t=0}=0,\quad \frac{d\Delta y_s}{d\theta}|_{t=0}=\frac{3s}{2}\sqrt{\frac{GM}{R}};
\end{align}
where the magnitude of the unpolarized electron's momentum has been set to $p=\sqrt{GM/R}$. Notably, in this work, the electron mass has been normalized to $m_e=1$.

\begin{figure}
  \centering
   \subcaptionbox{The difference in radial distances $\Delta |\vec{r}|$ }{
   \includegraphics[scale=0.44]{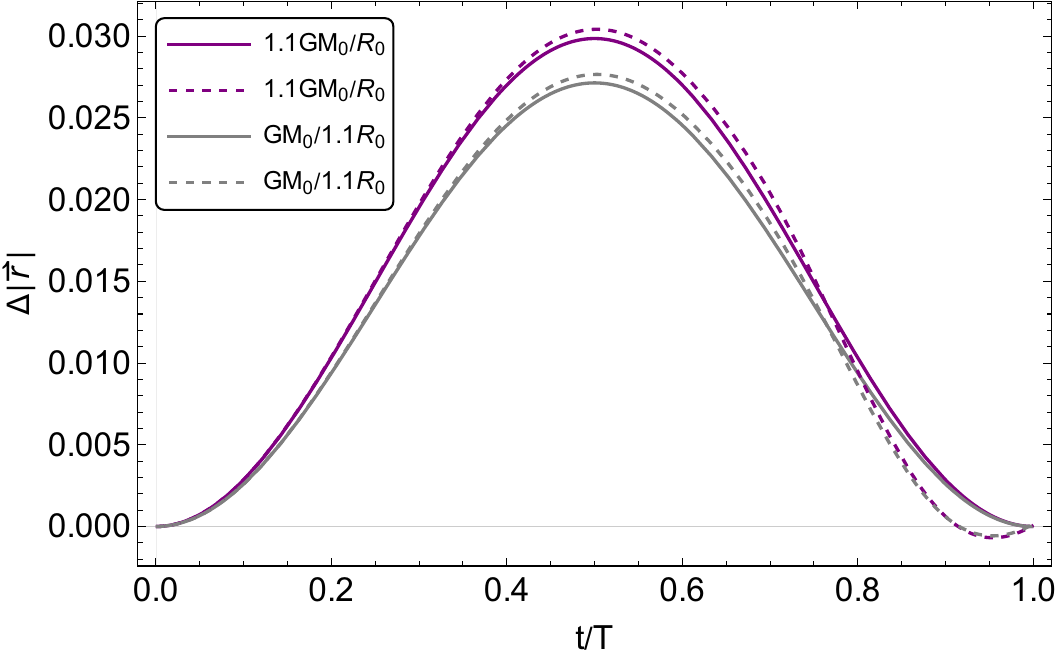}}
  \subcaptionbox{The separation between the electrons $|\Delta \vec{r}|$}{
   \includegraphics[scale=0.42]{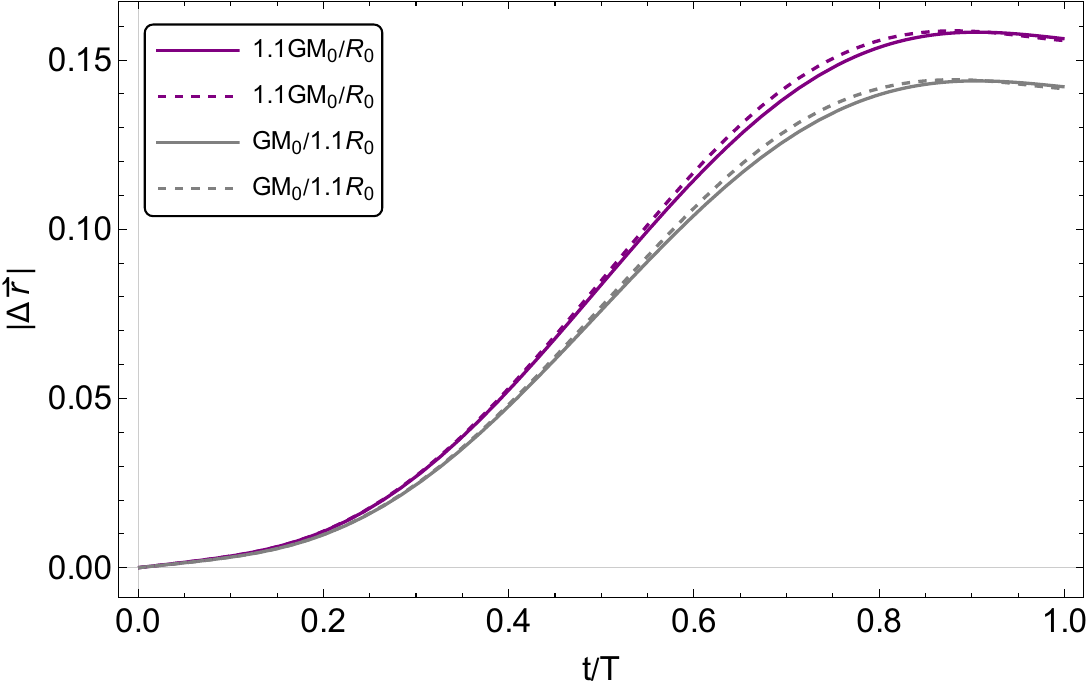}}
   \caption{Results given by the semi-classical approximations. The left panel illustrates the difference in radial distances, denoted as $\Delta |\vec{r}|$, from the gravitational source for two electrons with opposite spins $s=\pm 1/2$. The right panel shows the separation between these electrons, represented as $|\Delta \vec{r}|$. Dashed lines in both panels indicate numerical results calculated from the non-relativistic FW Hamiltonian $\mathcal{H}_{FW}$, with the parameters specified as $GM_0=1$ and $R_0=10^5$.}\label{fig3}
\end{figure}

By solving Eq. \eqref{c-sep} numerically, under the initial conditions detailed in Eq. \eqref{c-init}, we can calculate the trajectory differences for two electrons with opposite spins, $s = \pm 1/2$, as shown in Fig. \ref{fig3}. This figure illustrates that the difference in radial distances ($\Delta |\vec{r}|$) and spatial separation ($|\Delta \vec{r}|$) between these electrons, as predicted by the semi-classical equation \eqref{c-sep}, align closely with those calculated from the non-relativistic FW Hamiltonian ($\mathcal{H}_{FW}$).

For these electrons, the maximum difference in radial distance from the gravitational source ($\Delta |\vec{r}|_{\text{max}}$), as well as their spatial separations after half and a full orbital period ($|\Delta \vec{r}|_\text{T/2}$ and $|\Delta \vec{r}|_\text{T}$), can be approximated as:
\begin{align}\label{c-spat}
  \Delta |\vec{r}|_{\text{max}} \sim 9.0\sqrt{\frac{GM}{R}}, \quad |\Delta \vec{r}|_\text{T/2} \sim 8.1\pi\sqrt{\frac{GM}{R}}, \quad |\Delta \vec{r}|_\text{T} \sim 15.0\pi\sqrt{\frac{GM}{R}}.
\end{align}
Table \ref{tab1} details these quantities given by the numerical solutions of quantum Hamiltonian $\mathcal{H}_{FW}$. As indicated in this table, the values for $|\vec{r}|_{\text{max}}$ and $|\Delta \vec{r}|_\text{T/2}$ closely match the predictions made by the above equations.

\begin{table}
  \centering
  \setlength{\tabcolsep}{1.5mm}{
  \begin{tabular}{|cccc|}
  \toprule[0.08em]
  $GM/R$& $\Delta |\vec{r}|_{\text{max}}$& $|\Delta \vec{r}|_\text{T/2}$ & $|\Delta \vec{r}|_\text{T}$\\
  \midrule[0.05em]
  $1.3GM_0/R_0$&0.0331 $\textbf{(0.0324)}$&0.092 $\textbf{(0.092)}$ &0.174 $\textbf{(0.170)}$\\
  $1.2GM_0/R_0$&0.0318 $\textbf{(0.0312)}$&0.088 $\textbf{(0.088)}$&0.166  $\textbf{(0.163)}$\\
  $1.1GM_0/R_0$&0.0304 $\textbf{(0.0298)}$&0.085 $\textbf{(0.084)}$&0.157 $\textbf{(0.156)}$\\
  $1.0GM_0/R_0$&0.0290 $\textbf{(0.0285)}$&0.081 $\textbf{(0.080)}$&0.149 $\textbf{(0.149)}$\\
  $GM_0/1.1R_0$ & 0.0276 $\textbf{(0.0271)}$&0.077 $\textbf{(0.077)}$& 0.141 $\textbf{(0.142)}$\\
  $GM_0/1.2R_0$ & 0.0264 $\textbf{(0.0260)}$&0.074 $\textbf{(0.073)}$& 0.136 $\textbf{(0.136)}$\\
  $GM_0/1.3R_0$ & 0.0254 $\textbf{(0.0250)}$&0.071 $\textbf{(0.071)}$& 0.130 $\textbf{(0.131)}$\\
  \bottomrule[0.08em]
  \end{tabular}}
  \caption{Results as predicted by Schrödinger equation versus semi-classical approximations. The values enclosed in bold parentheses, "$\textbf{(\quad)}$", correspond to those results calculated using Eq. \eqref{c-spat}.}
  \label{tab1}
  \end{table}

Although minor deviations from Eq. \eqref{c-spat} in the spatial separation $|\Delta \vec{r}|$ are observed after a full orbital period, these deviations are generally modest, typically under $10\%$ of the respective spatial separations, and in certain cases, as minimal as $2\%$. As discussed in Sec. \ref{sec.sep}, such deviations are likely due to numerical errors, affirming that the spatial separation after a full orbital period can still be approximated as $|\Delta \vec{r}|_T \sim 15.0\pi\sqrt{GM/R}$.

Therefore, under the non-relativistic approximation, the motion of electrons can be effectively described by the semi-classical Hamiltonian $H_{\text{class}}$ within the weak Schwarzschild gravitational field.  This suggests that, when dealing with electrons moving in weak gravitational fields, such as those of Earth or the Sun, detailed numerical solutions of the Schrödinger equation might not be critically urgent.

\begin{figure}
  \centering
   \includegraphics[scale=0.7]{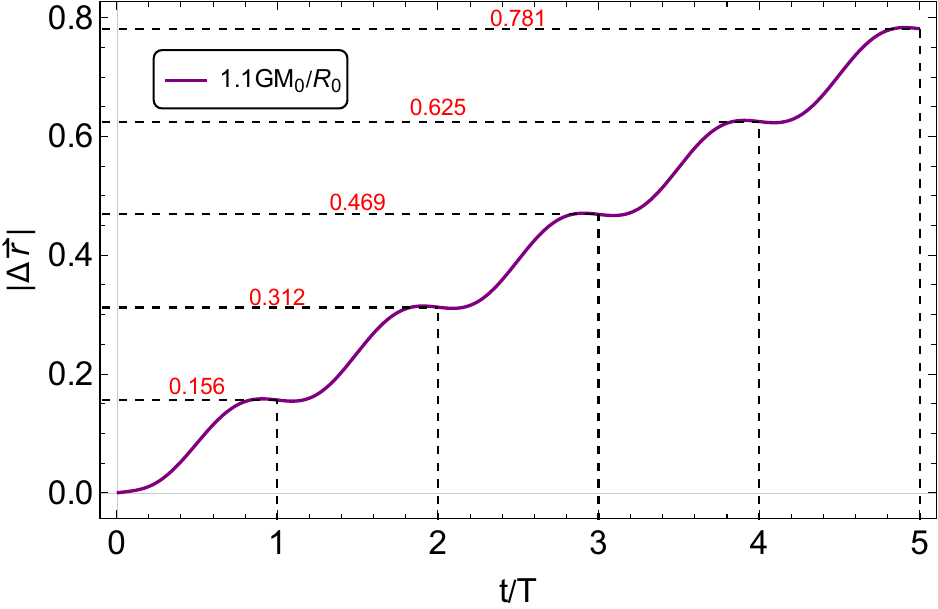}
   \caption{Separation between these two electrons vs. the number of orbital periods. Here, the parameters are specified as $GM_0=1$ and $R_0=10^5$.}\label{fig4}
\end{figure}

The relationship between the separation $|\Delta \vec{r}|$ and the number of orbital periods is shown in Fig. \ref{fig4}. It is observed that the spatial separation increases linearly with the number of orbital periods as given by
\begin{equation}
|\Delta \vec{r}|_\text{n}\simeq n |\Delta \vec{r}|_\text{T}\sim 15.0n\pi\sqrt{\frac{GM}{R}},
\end{equation}
where $n$ represents the number of orbital periods. This linear relationship, consistent with Eq. \eqref{n-orbit} derived from the quantum Hamiltonian $\mathcal{H}_{FW}$, confirms that the spatial separation between two electrons with opposite spins is directly proportional to the number of orbital periods. Such proportionality significantly enhances the detectability of $|\Delta \vec{r}|$, thereby increasing the potential of observing the gravitational SHE in experiments.

\section{Discussion}\label{sec.Di}

In this study, we investigate the dynamics of electrons by numerically solving the Schrödinger equation with non-relativistic FW Hamiltonian $\mathcal{H}_{FW}$, in the weak Schwarzschild gravitational field. Our findings indicate that two electrons with opposite spins, $s = \pm 1/2$, are expected to follow different trajectories when their orbital plane is perpendicular to their spins, a result of gravitational spin-orbit coupling, known as the gravitational spin Hall effect (SHE). In this work, we also apply a semi-classical approximation by substituting quantum operators in the Hamiltonian $\mathcal{H}_{FW}$ with their expectation values. Within the weak field approximation ($GM/r\ll 1$), our findings demonstrate that this semi-classical approach effectively captures the motion of spin-polarized electrons, thereby streamlining the analysis of electron dynamics in gravitational fields.

In the gravitational fields, the momentum-velocity relationship for a spin-polarized electron is expected to be altered by gravitational spin-orbit coupling, deviating from the classical $\vec{p}=m_e\vec{v}$ relation under the non-relativistic approximation. Our findings indicate that electrons with opposite spins, $s=\pm 1/2$, launched from the same point $\vec{r}_0=(R,0,0)$ with identical initial momentum $\vec{p}_0=(0,\sqrt{GM/R},0)$, are predicted to trace distinct orbits around the gravitational source, characterized by varying eccentricities. Consequently, these electrons are expected to separate from each other, with their separation increasing over subsequent orbital periods.

More specifically, the separation between the electrons after $n$ orbital periods can be approximated as $|\Delta \vec{r}|_\text{n}\simeq n|\Delta \vec{r}|_\text{T}$, where $|\Delta \vec{r}|_\text{T}\sim 15.0\pi\sqrt{GM/R}/m_e$ represents their separation after one complete orbit. Thus, as the electrons continue to orbit the gravitational source, their separation $|\Delta \vec{r}|$ grows linearly with the number of orbital periods. This linear progression can greatly enchance the detectability of $|\Delta \vec{r}|$, improving the feasibility of observing the gravitational SHE in experiments.

To illustrate, consider the two electrons orbiting the Earth at a low orbit radius of approximately $R_\oplus \sim 6.4 \times 10^6\, \text{m}$.  The gravitational potential is approximated as $GM_\oplus/R_\oplus \simeq 7.0 \times 10^{-10}$.  For these two electrons  with opposite spin $s=\pm 1/2$, their separation after a full orbital period ($T \simeq 2\pi R_\oplus / |\vec{p}_0| \sim 84\, \text{min}$), is approximated as $|\Delta \vec{r}|_\text{T} \sim 4.7 \times 10^{-16}\, \text{m}$. Over a year, equal to about 6257 orbital periods,  the separation approximately increases to  $|\Delta \vec{r}| \simeq 6257 |\Delta \vec{r}|_\text{T} \sim 3.0 \times 10^{-12}\, \text{m}$. Compared to the initial separation $|\Delta \vec{r}|_\text{T}$, the separation $|\Delta \vec{r}|$ between the two electrons significantly increases over the course of a year. Consequently, for spin-polarized electrons on artificial satellites, their separation becomes theoretically more detectable, despite technological challenges.

\begin{acknowledgments}
  This work is  supported by the National Natural Science Foundation of China (Grant No. 12375084).
\end{acknowledgments}

\bibliographystyle{apsrev4-2}
\bibliography{reference.bib}

\end{document}